\documentclass[aps,prl,twocolumn,showpacs,superscriptaddress]{revtex4}
\usepackage{graphicx}
\usepackage{dcolumn}
\usepackage{bm}
\usepackage{amsmath}
\usepackage{amssymb}

\begin{document}

\preprint{\vbox{ \hbox{Belle Preprint 2007-xx}
                 \hbox{KEK   Preprint 2007-xx}
                 \hbox{Version 3.7 \today}
                              }}

\title{\quad\\
Observation of anomalous $\Upsilon(1S)\pi^+ \pi^-$ and $\Upsilon(2S)\pi^+ \pi^-$
production near the $\Upsilon(5S)$ resonance}

\begin{abstract}
We report the first observation of $e^+e^- \to
\Upsilon(1S)\pi^+\pi^-$, $\Upsilon(2S)\pi^+\pi^-$, and first
evidence for $e^+e^- \to \Upsilon(3S)\pi^+\pi^-$,
$\Upsilon(1S)K^+K^-$, near the peak of the $\Upsilon(5S)$
resonance at $\sqrt{s}\sim10.87$ GeV. The results are based on a
data sample of 21.7~fb$^{-1}$ collected with the Belle detector at
the KEKB $e^+e^-$ collider. 
Attributing the signals to the $\Upsilon(5S)$ resonance, 
the partial widths
$\Gamma(\Upsilon(5S)\to\Upsilon(1S)\pi^+\pi^-) = 0.59\pm0.04{\rm (stat)}\pm0.09{\rm (syst)}$ MeV and
$\Gamma(\Upsilon(5S)\to\Upsilon(2S)\pi^+\pi^-) = 0.85\pm0.07{\rm (stat)}\pm0.16{\rm (syst)}$ MeV 
are obtained from the observed cross sections. 
These values exceed by more than two orders of magnitude 
the previously measured partial widths for dipion transitions between lower $\Upsilon$ resonances. 
\end{abstract}


\affiliation{Budker Institute of Nuclear Physics, Novosibirsk}
\affiliation{Chiba University, Chiba}
\affiliation{University of Cincinnati, Cincinnati, Ohio 45221}
\affiliation{Justus-Liebig-Universit\"at Gie\ss{}en, Gie\ss{}en}
\affiliation{The Graduate University for Advanced Studies, Hayama}
\affiliation{Hanyang University, Seoul}
\affiliation{University of Hawaii, Honolulu, Hawaii 96822}
\affiliation{High Energy Accelerator Research Organization (KEK), Tsukuba}
\affiliation{Institute of High Energy Physics, Chinese Academy of Sciences, Beijing}
\affiliation{Institute of High Energy Physics, Vienna}
\affiliation{Institute of High Energy Physics, Protvino}
\affiliation{Institute for Theoretical and Experimental Physics, Moscow}
\affiliation{J. Stefan Institute, Ljubljana}
\affiliation{Kanagawa University, Yokohama}
\affiliation{Korea University, Seoul}
\affiliation{Kyungpook National University, Taegu}
\affiliation{\'Ecole Polytechnique F\'ed\'erale de Lausanne (EPFL), Lausanne}
\affiliation{Faculty of Mathematics and Physics, University of Ljubljana, Ljubljana}
\affiliation{University of Maribor, Maribor}
\affiliation{University of Melbourne, School of Physics, Victoria 3010}
\affiliation{Nagoya University, Nagoya}
\affiliation{Nara Women's University, Nara}
\affiliation{National Central University, Chung-li}
\affiliation{National United University, Miao Li}
\affiliation{Department of Physics, National Taiwan University, Taipei}
\affiliation{H. Niewodniczanski Institute of Nuclear Physics, Krakow}
\affiliation{Nippon Dental University, Niigata}
\affiliation{Niigata University, Niigata}
\affiliation{University of Nova Gorica, Nova Gorica}
\affiliation{Osaka City University, Osaka}
\affiliation{Osaka University, Osaka}
\affiliation{Panjab University, Chandigarh}
\affiliation{Saga University, Saga}
\affiliation{University of Science and Technology of China, Hefei}
\affiliation{Seoul National University, Seoul}
\affiliation{Sungkyunkwan University, Suwon}
\affiliation{University of Sydney, Sydney, New South Wales}
\affiliation{Toho University, Funabashi}
\affiliation{Tohoku Gakuin University, Tagajo}
\affiliation{Department of Physics, University of Tokyo, Tokyo}
\affiliation{Tokyo Institute of Technology, Tokyo}
\affiliation{Tokyo Metropolitan University, Tokyo}
\affiliation{Tokyo University of Agriculture and Technology, Tokyo}
\affiliation{Virginia Polytechnic Institute and State University, Blacksburg, Virginia 24061}
\affiliation{Yonsei University, Seoul}

 \author{K.-F.~Chen}\affiliation{Department of Physics, National Taiwan University, Taipei} 
 \author{W.-S.~Hou}\affiliation{Department of Physics, National Taiwan University, Taipei} 
 \author{M.~Shapkin}\affiliation{Institute of High Energy Physics, Protvino} 
 \author{A.~Sokolov}\affiliation{Institute of High Energy Physics, Protvino} 
  \author{I.~Adachi}\affiliation{High Energy Accelerator Research Organization (KEK), Tsukuba} 
  \author{H.~Aihara}\affiliation{Department of Physics, University of Tokyo, Tokyo} 
  \author{K.~Arinstein}\affiliation{Budker Institute of Nuclear Physics, Novosibirsk} 
  \author{V.~Aulchenko}\affiliation{Budker Institute of Nuclear Physics, Novosibirsk} 
  \author{T.~Aushev}\affiliation{\'Ecole Polytechnique F\'ed\'erale de Lausanne (EPFL), Lausanne}\affiliation{Institute for Theoretical and Experimental Physics, Moscow} 
  \author{A.~M.~Bakich}\affiliation{University of Sydney, Sydney, New South Wales} 
% ...
\collaboration{The Belle Collaboration}

\maketitle

\tighten

Heavy quarkonia provide a unique nonrelativistic system in which low energy QCD may be illuminated
through their energy levels, widths, and transition amplitudes. 
Dipion transitions between $\psi$ and $\Upsilon$ levels below the open 
flavor thresholds have been successfully
described in terms of multipole moments of the QCD field~\cite{Brown:1975dz}. 
The first measurements above the open beauty
threshold, namely of $\Upsilon(4S) \to \Upsilon(1S)\pi^+\pi^-$~\cite{CLEO:2007sja,Aubert:2006bm,Sokolov:2006sd}, 
are consistent with this picture~\cite{Simonov:2007bm}. 
(The $\Upsilon(4S)$ is the third radial excitation of the $J^{PC} = 1^{--}$ state $\Upsilon(1S)$.)

The spectroscopy above open flavor threshold is complex, 
however, as there is no positronium analogue. The
recent discovery of a broad $1^{--}$ state, the $Y(4260)$,
decaying with an unexpectedly large partial width to
$J/\psi \pi^+\pi^-$~\cite{Y4260}, has brought new challenges to the interpretation of its composition, 
with ``hybrid" $c\overline{c}g$ (where $g$
is a gluon) and $c\overline{c}q\overline{q}$ (where $q\overline{q}$ is a color-octet light quark
pair) four quark state as possibilities. The observation of
a bottomonium counterpart to $Y(4260)$, which we shall
refer to as $Y_b$~\cite{Hou:2006it}, could shed further light on the structure
of such particles. The expected mass is above the $\Upsilon(4S)$.
It has been suggested that a $Y_b$ with lower mass can be
searched for by radiative return from the $\Upsilon(5S)$, and one
with higher mass through an anomalous rate of $\Upsilon(nS)\pi\pi$
events~\cite{Hou:2006it}; scaling from $\Upsilon(4S)\to\Upsilon(1S)\pi\pi$, one expects
$\Upsilon(5S)\to\Upsilon(1S)\pi\pi$ to have branching fraction $\sim 10^{-5}$.

Here we report the first observation of
$\Upsilon(1S)\pi^+\pi^-$ and $\Upsilon(2S)\pi^+\pi^-$ final
states, as well as evidence for $\Upsilon(3S)\pi^+\pi^-$ and
$\Upsilon(1S)K^+K^-$ in a 21.7~fb$^{-1}$ data sample collected
near the peak of the $\Upsilon(5S)$ resonance with the Belle
detector at the KEKB $e^+e^-$ energy-asymmetric collider~\cite{ref:KEKB}. 
The rates for $\Upsilon(1S) \pi^+ \pi^-$ and $\Upsilon(2S) \pi^+ \pi^-$
are much larger than the expectations from scaling the
comparable $\Upsilon(4S)$ decays to the $\Upsilon(5S)$.
Since only one center-of-mass (CM) energy is
used, one does not know whether 
these enhancements are an effect of
the $\Upsilon(5S)$ itself, or due to a nearby or overlapping $``Y_b"$ state. 
Throughout this Letter, we shall
therefore use the notation $\Upsilon(10860)$ instead of
$\Upsilon(5S)$.

The Belle detector is a large-solid-angle magnetic spectrometer, 
which consists of a silicon vertex detector
(SVD), a central drift chamber (CDC), an array of
aerogel threshold Cherenkov counters (ACC), a barrel-like arrangement of
time-of-flight scintillation counters (TOF), and an electromagnetic
calorimeter comprised of CsI(Tl) crystals (ECL) located inside a
superconducting solenoid that provides a 1.5~T magnetic field. An iron
flux-return located outside the coil is instrumented to detect $K_L^0$
mesons and to identify muons (KLM). The detector is described in detail
elsewhere~\cite{ref:belle_detector}.

The $\Upsilon(10860)\to\Upsilon(nS)\pi^+\pi^-$ and
$\Upsilon(1S)K^+K^-$ final states are reconstructed using
$\Upsilon(nS) \to \mu^+\mu^-$ decays. 
Events with exactly four well-constrained charged tracks
and zero net charge are selected.
Muon candidates are required
to have hits in the KLM detector associated with the extrapolated 
trajectory of
the charged track. Two muons with opposite charge are selected to
form a $\Upsilon(nS)$ candidate.
The two remaining tracks are treated as pion or kaon candidates. To
suppress the background from $\mu^+\mu^-\gamma \to
\mu^+\mu^-e^+e^-$ with photon conversion, pion candidates with
positive electron identification are rejected. Electron
identification is based on associating the ECL shower energy to the track
momentum, $dE/dx$ from CDC, and the ACC response. 
Kaon candidates are required to have a kaon likelihood,
estimated with information from the ACC,
TOF, and $dE/dx$ from the CDC, greater than 0.1. This requirement has an
efficiency of 98.2\%. 
The cosine of the
opening angle between the $\pi^+$ and $\pi^-$ ($K^+$ and $K^-$)
momenta in the laboratory frame is required to be less than 0.95.
The trigger efficiency is found to be very
close to 100\% for these final states.
%
%
%
To reject (radiative) Bhabha and $\mu$-pair backgrounds, the data are 
required to satisfy either $\theta_{\rm max}<175^\circ$, or 2 GeV $<\sum
E_{\rm ECL} < 10$ GeV, where $\theta_{\rm max}$ is the 
maximum opening angle between any charged tracks in the CM frame,
and $\sum E_{\rm ECL}$ is the sum of the ECL clusters' energy.

The signal candidates are identified using the kinematic variable
$\Delta M$, defined as the difference between
$M(\mu^+\mu^-\pi^+\pi^-)$ or $M(\mu^+\mu^-K^+K^-)$ and
$M(\mu^+\mu^-)$ for pion or kaon modes. Sharp peaks are expected at
$\Delta M = M_{\Upsilon(mS)} - M_{\Upsilon(nS)}$ for $m>n$. For
$\Upsilon(10860)\to\Upsilon(nS)\pi^+\pi^-$ and
$\Upsilon(1S)K^+K^-$, signal events should be concentrated at $\Delta
M = \sqrt{s} - M_{\Upsilon(nS)}$, since a single CM energy is
used. 

\begin{figure}[b!]
\begin{center}
\includegraphics[width=8.5cm]{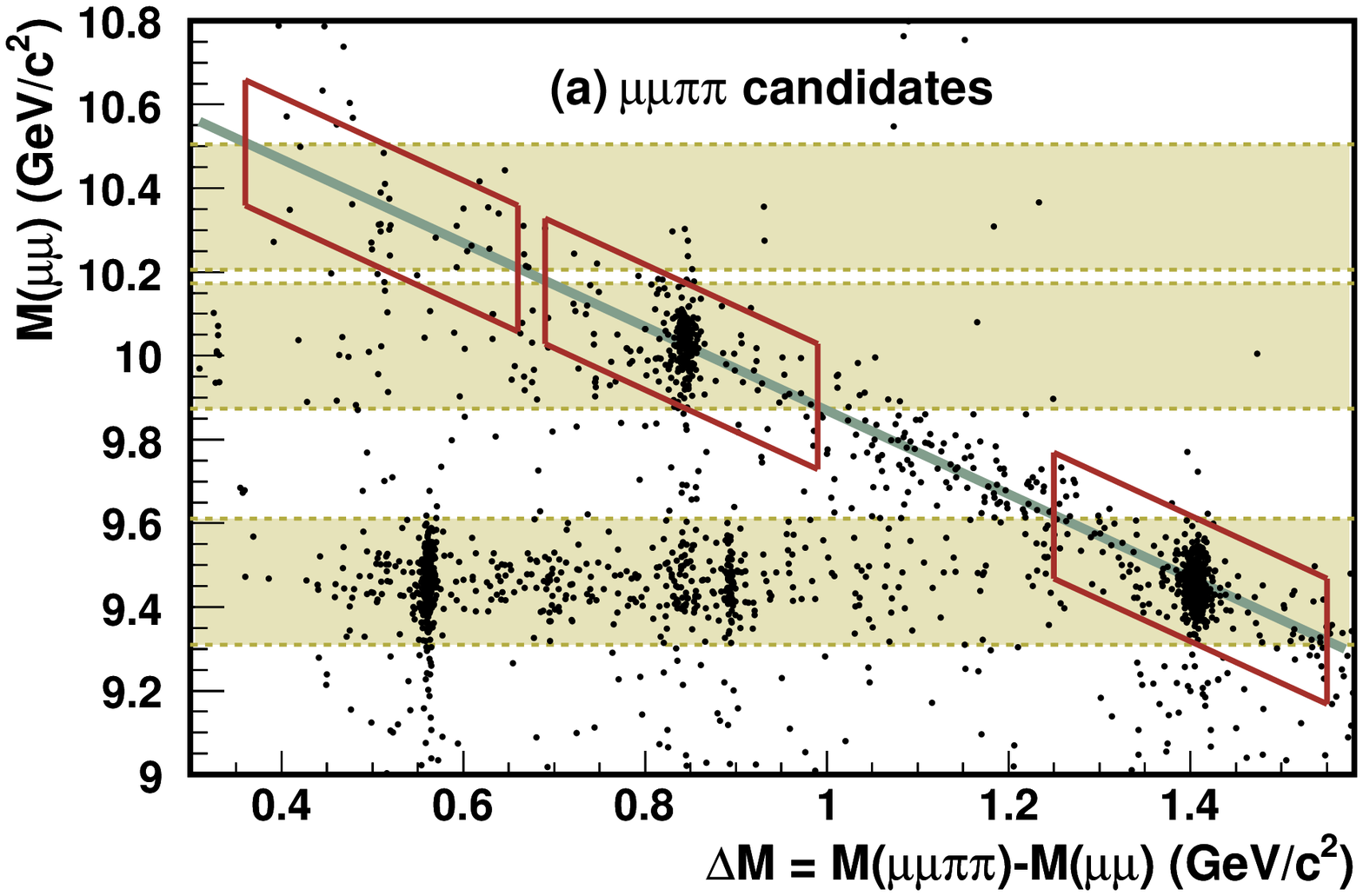}
\includegraphics[width=8.5cm]{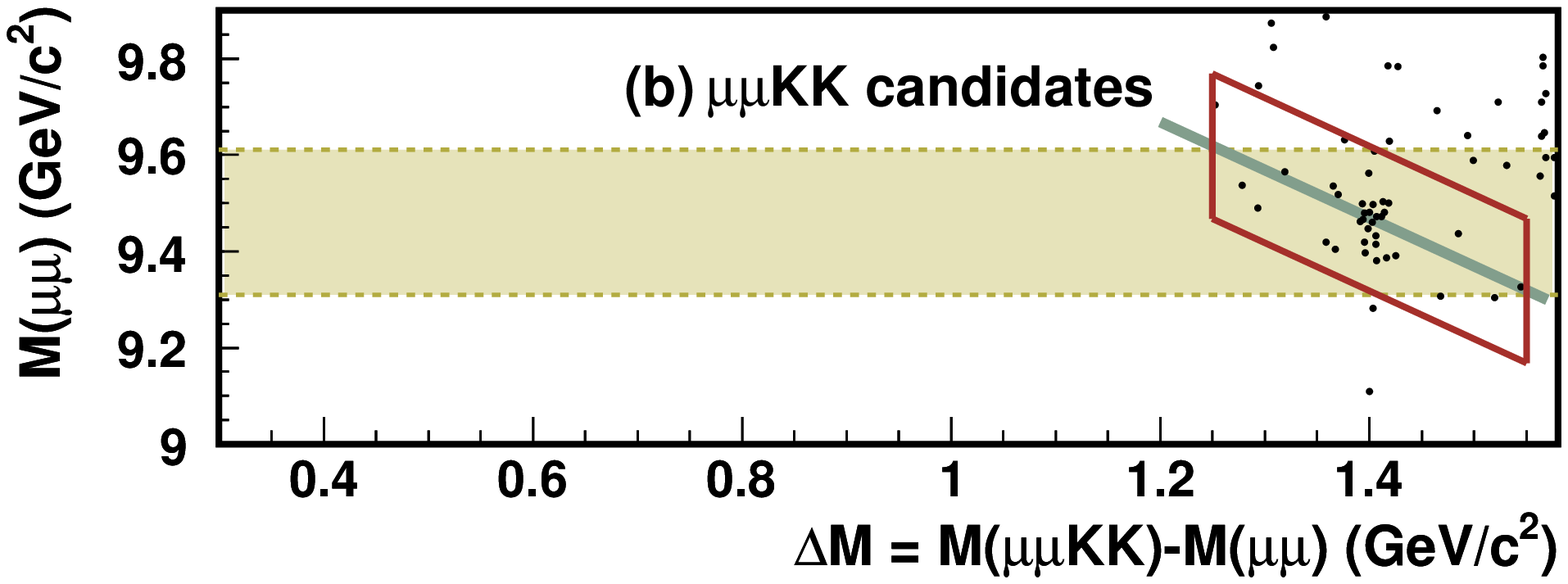}
\end{center}
\caption{Scatter plot of $M(\mu^+\mu^-)$ vs.
$\Delta M$ for the data collected at $\sqrt{s}\sim10.87$ GeV, for
(a) $\mu^+\mu^-\pi^+\pi^-$ and (b) $\mu^+\mu^-K^+K^-$ candidates.
Horizontal shaded bands correspond to $\Upsilon(1S)$,
$\Upsilon(2S)$ and $\Upsilon(3S)$ (only $\Upsilon(1S)$ for (b)),
and open boxes are the fitting regions for
$\Upsilon(10860)\to\Upsilon(nS)\pi^+\pi^-$ and
$\Upsilon(1S)K^+K^-$. The lines indicate the kinematic boundaries, 
$M(\mu^+\mu^-\pi^+\pi^-, \mu^+\mu^-K^+K^-) = \sqrt{s}$.}
 \label{fig:ym_vs_dm}
\end{figure}

Figure~\ref{fig:ym_vs_dm} shows the
two-dimensional scatter plot of $M(\mu^+\mu^-)$ vs. $\Delta M$ for
the data. Clear enhancements are observed, especially for
$\Upsilon(10860)\to\Upsilon(1S)\pi^+\pi^-$ and
$\Upsilon(2S)\pi^+\pi^-$ decays. 
The dominant background processes, $e^+e^- \to \mu^+\mu^-\gamma(\to e^+e^-)$ and $e^+e^- \to \mu^+\mu^-\pi^+\pi^-$ 
accumulate at the kinematic boundary, $M(\mu^+\mu^-\pi^+\pi^-) = \sqrt{s}$.
The events with $|M(\mu^+\mu^-\pi^+\pi^-)-\sqrt{s}|<150$ MeV
or $|M(\mu^+\mu^-K^+K^-)-\sqrt{s}|<150$ MeV are selected. The
fitting regions are defined by 1.25 GeV/$c^2$ $<\Delta M<$ 1.55
GeV/$c^2$, 0.69 GeV/$c^2$ $<\Delta M<$ 0.99 GeV/$c^2$, and 0.36
GeV/$c^2$ $<\Delta M<$ 0.66 GeV/$c^2$ for
$\Upsilon(10860)\to\Upsilon(1S)\pi^+\pi^-$,
$\Upsilon(2S)\pi^+\pi^-$, and $\Upsilon(3S)\pi^+\pi^-$,
respectively. The fitting region in $\Delta M$ for
$\Upsilon(10860)\to\Upsilon(1S)K^+K^-$ is the same as for the
$\Upsilon(1S)\pi^+\pi^-$ mode. 
The oblique fitting regions 
are selected so that the background shape is monotonic along each band.
The background distributions are verified 
using the off-resonance sample (recorded at $\sqrt{s} \sim 10.52$ GeV)~\cite{Sokolov:2006sd}.

\begin{figure}[b!]
\begin{center}
\includegraphics[width=8.5cm]{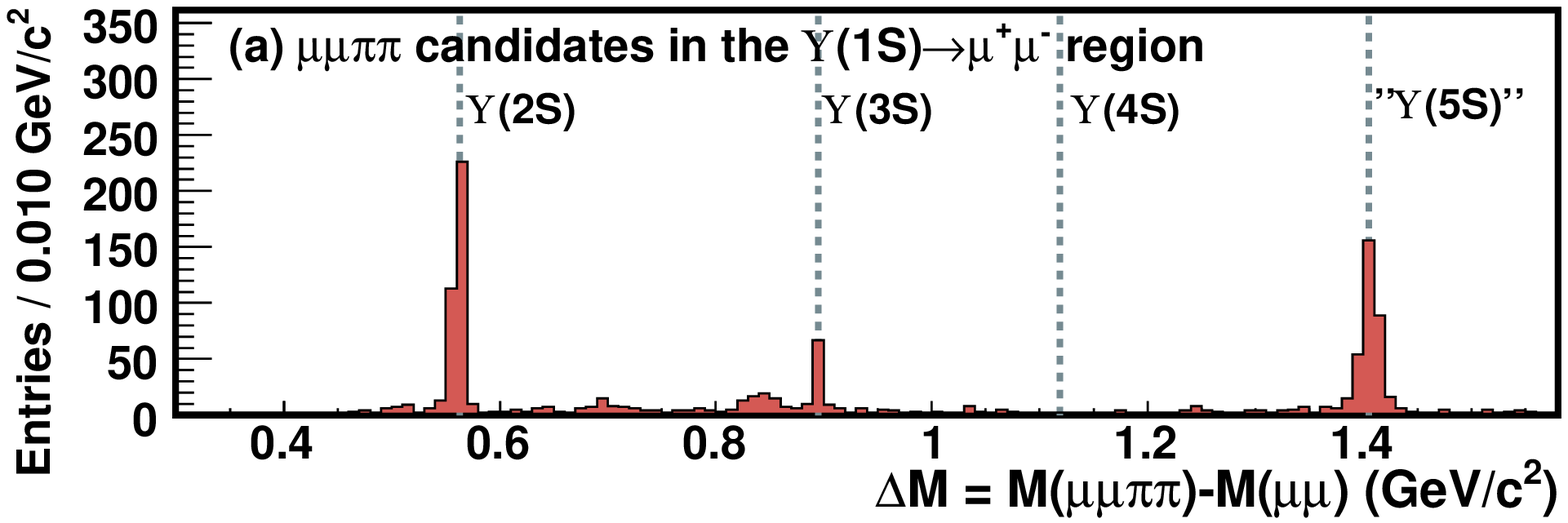}
\includegraphics[width=8.5cm]{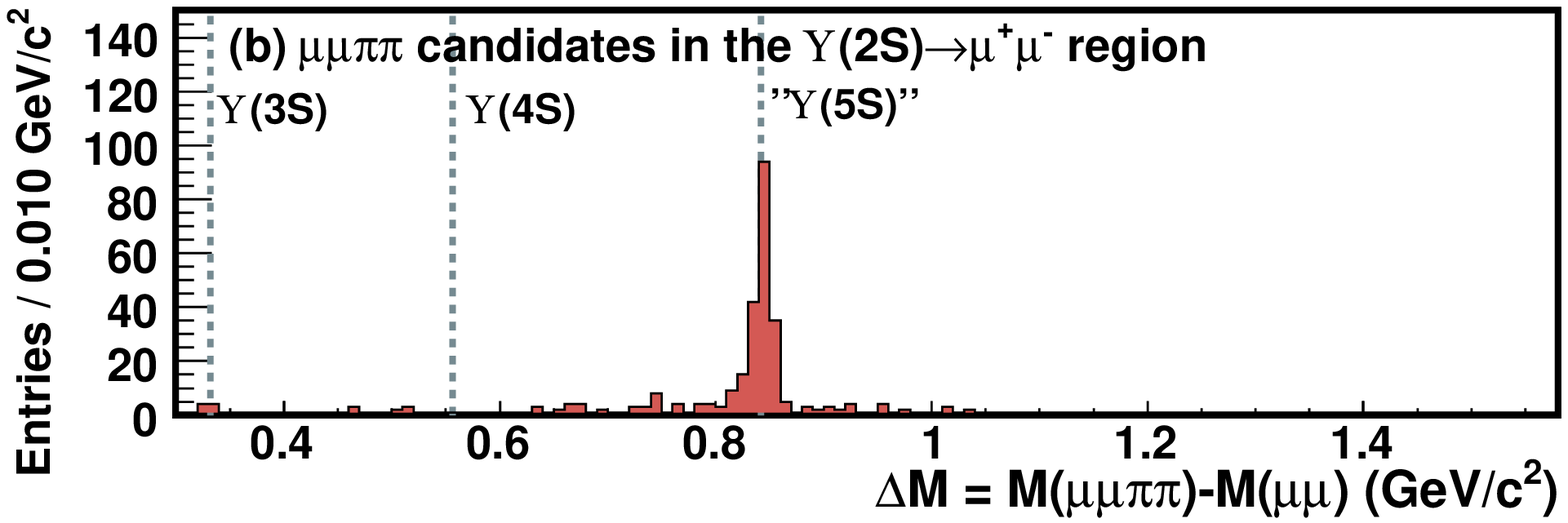}
\end{center}
\caption{The $\Delta M$ distributions for $\mu^+\mu^-\pi^+\pi^-$
events in the (a) $\Upsilon(1S)\to \mu^+\mu^-$ and (b)
$\Upsilon(2S)\to \mu^+\mu^-$ bands of Fig.~\ref{fig:ym_vs_dm}(a).
Vertical dashed lines show the expected $\Delta M$ values for 
the $\Upsilon(nS)\to\Upsilon(1,2S)\pi^+\pi^-$ transitions.}
 \label{fig:dm}
\end{figure}

The $\Delta M$
distributions for the $\mu^+\mu^-\pi^+\pi^-$ candidates in the
$\Upsilon(1S)$ and $\Upsilon(2S) \to \mu^+\mu^-$ mass bands are shown
in Fig.~\ref{fig:dm}. The peaks for
$\Upsilon(10860)\to\Upsilon(1S)\pi^+\pi^-$ and
$\Upsilon(2S)\pi^+\pi^-$ are located at $\Delta M
\sim 1.41$ GeV/$c^2$ and $\sim0.84$ GeV/$c^2$, respectively. Two other peaks at
$\Delta M \sim 0.56$ GeV/$c^2$ and $\sim0.89$ GeV/$c^2$ correspond
to $\Upsilon(2S)\to\Upsilon(1S)\pi^+\pi^-$ and
$\Upsilon(3S)\to\Upsilon(1S)\pi^+\pi^-$ transitions, respectively.
The absence of a peak around 1.12 GeV/$c^2$ corresponding to
$\Upsilon(4S)\to\Upsilon(1S)\pi^+\pi^-$ is consistent with
the rates measured in Refs.~\cite{Aubert:2006bm,Sokolov:2006sd}. 
The structure just below $\Upsilon(3S)\to\Upsilon(1S)\pi^+\pi^-$ 
in the $\Delta M$ distribution
is from the cascade decays 
$\Upsilon(10860)\to\Upsilon(2S)\pi^+\pi^-$ with $\Upsilon(2S) \to \Upsilon(1S)[\to \mu^+\mu^-]X$.

\begin{figure}[t!]
\begin{center}
\includegraphics[width=4.25cm]{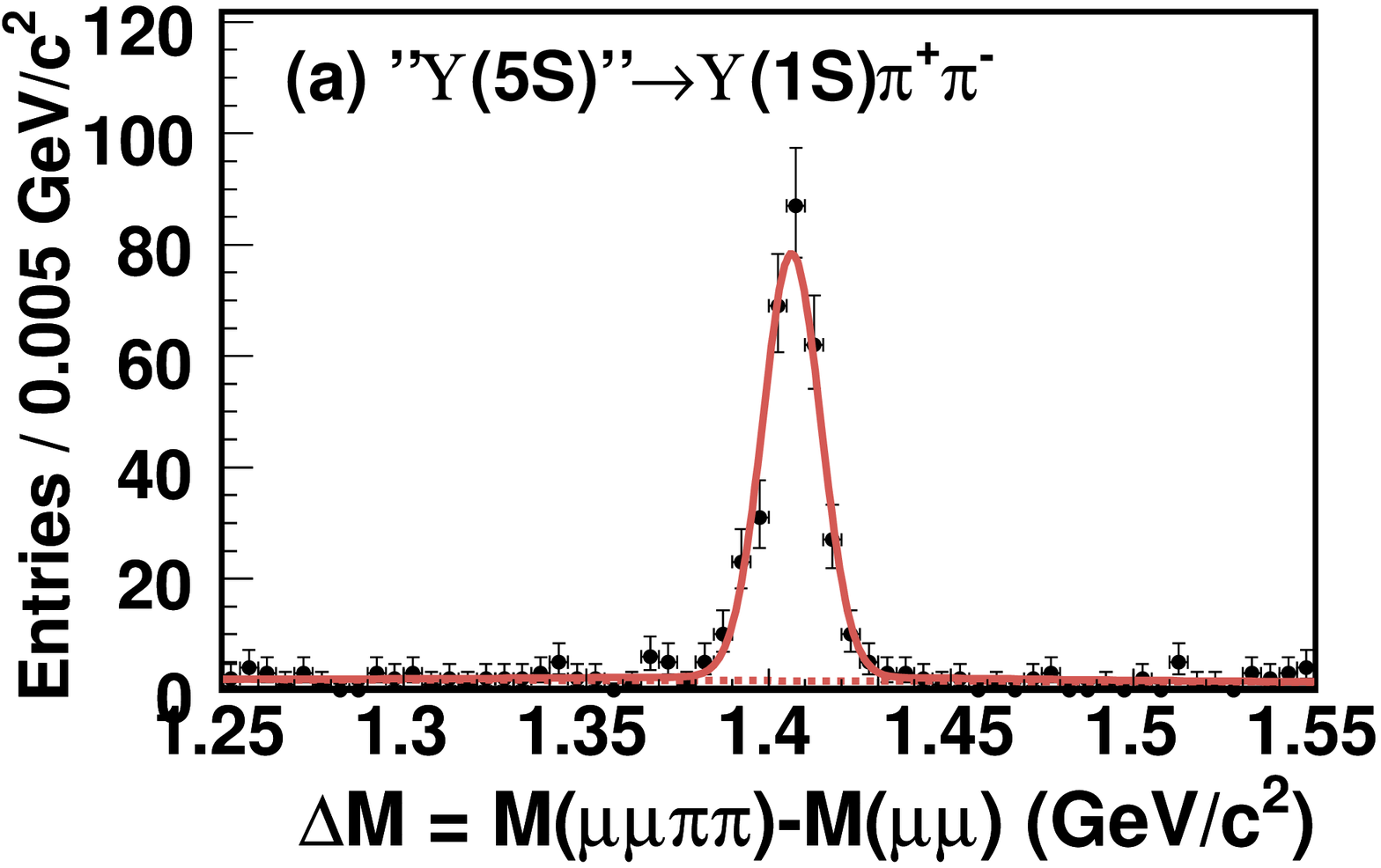}
\includegraphics[width=4.25cm]{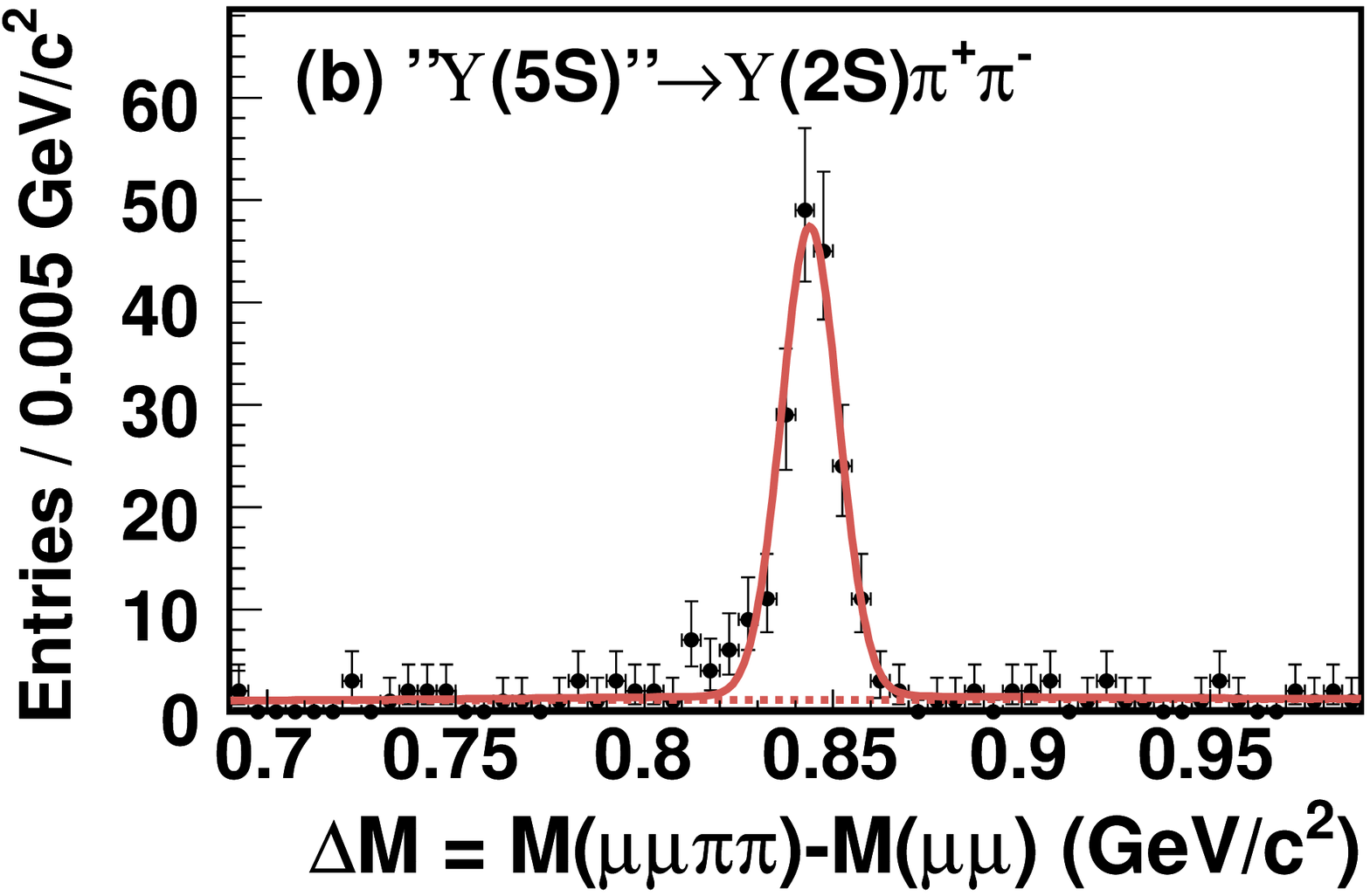}
\includegraphics[width=4.25cm]{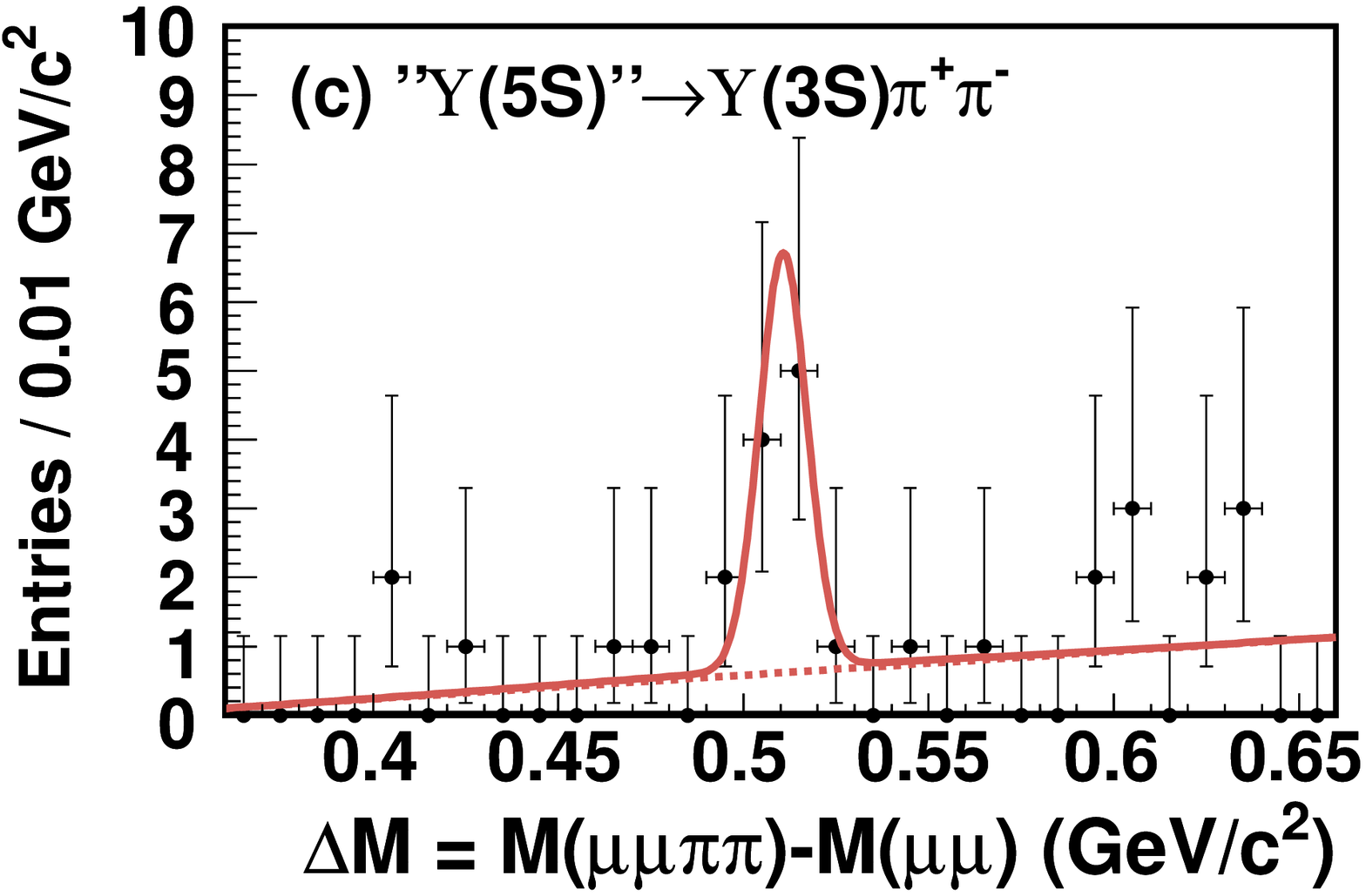}
\includegraphics[width=4.25cm]{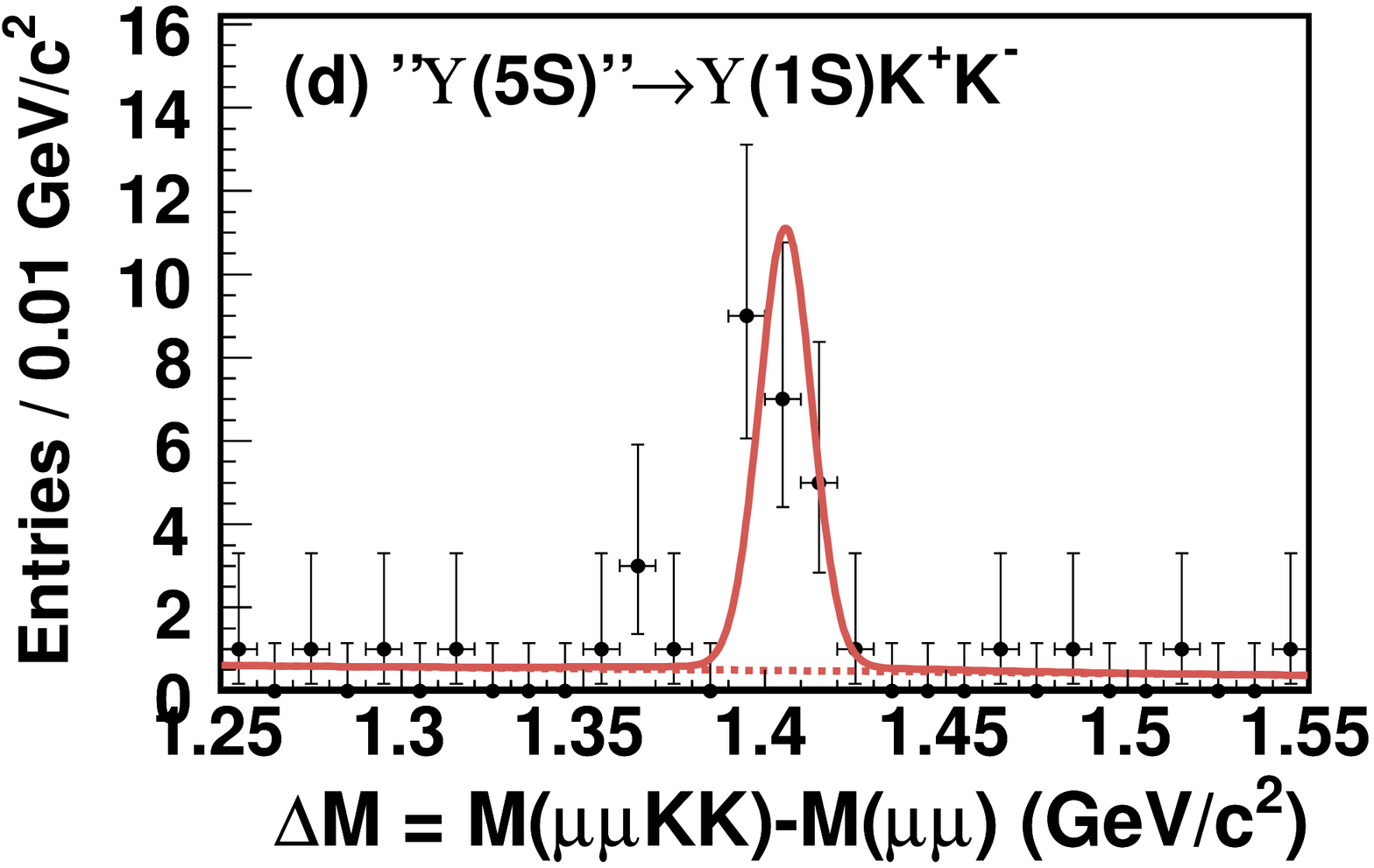}
\end{center}
\caption{The $\Delta M$ distributions for
 (a) $\Upsilon(1S)\pi^+\pi^-$,
 (b) $\Upsilon(2S)\pi^+\pi^-$,
 (c) $\Upsilon(3S)\pi^+\pi^-$, and
 (d) $\Upsilon(1S)K^+K^-$ with the fit results superimposed.
 The dashed curves show the background components in the fits.}
 \label{fig:dmfit}
\end{figure}

Signal yields are extracted by unbinned extended maximum likelihood (ML) fits to
the $\Delta M$ distributions. The likelihood for the fit is written as
\begin{equation}
\mathcal{L}(N_s,N_b) = {e^{-(N_s+N_b)} \over {N!}}
 \prod_{i=1}^N [ N_s \cdot P_s (\Delta M_i) + N_b \cdot P_b (\Delta M_i) ]~,
\end{equation}
where $N_s$ ($N_b$) denotes the yield for signal (background), and
$P_s$ ($P_b$) is the signal (background) probability density
function (PDF). 
The signal is described by a sum of two Gaussians 
while the background is approximated by a linear function.
The tail part of the signal PDF is parameterized by a 
broad Gaussian, whose
width and fraction (of area) are fixed from Monte
Carlo (MC) simulation.
For the $\Upsilon(10860)\to \Upsilon(1S)\pi^+\pi^-$ and 
$\Upsilon(2S)\pi^+\pi^-$ modes, the remaining PDF parameters 
and yields of signal and background are floated in the fits.
For the $\Upsilon(10860)\to \Upsilon(3S)\pi^+\pi^-$ and 
$\Upsilon(1S)K^+K^-$ transitions, where statistics are limited, 
the means and widths are established based on 
$\Upsilon(10860)\to \Upsilon(1S)\pi^+\pi^-$ events and 
fixed in the fits. 
We observe $325^{+20}_{-19}$, $186\pm15$, $10.5^{+4.0}_{-3.3}$,
and $20.2^{+5.2}_{-4.5}$ events in the
$\Upsilon(10860)\to\Upsilon(1S)\pi^+\pi^-$,
$\Upsilon(2S)\pi^+\pi^-$, $\Upsilon(3S)\pi^+\pi^-$, and
$\Upsilon(1S)K^+K^-$ channels, with significances of 20$\sigma$,
14$\sigma$, 3.2$\sigma$, and 4.9$\sigma$, respectively. 
The significance
is calculated using the difference in likelihood values of the
best fit and of a null signal hypothesis including the effect of systematic uncertainties.
The Gaussian widths of the $\Upsilon(10860)\to\Upsilon(1S)\pi^+\pi^-$ and $\Upsilon(2S)\pi^+\pi^-$
peaks are found to be $8.0\pm0.5$ MeV/$c^2$ and $7.6\pm0.7$ MeV/$c^2$, respectively,
and are consistent with the MC predictions.
The distributions of $\Delta M$ with the fit results
superimposed are shown in Fig.~\ref{fig:dmfit}. 

\begin{figure}[t!]
\begin{center}
\includegraphics[width=4.25cm]{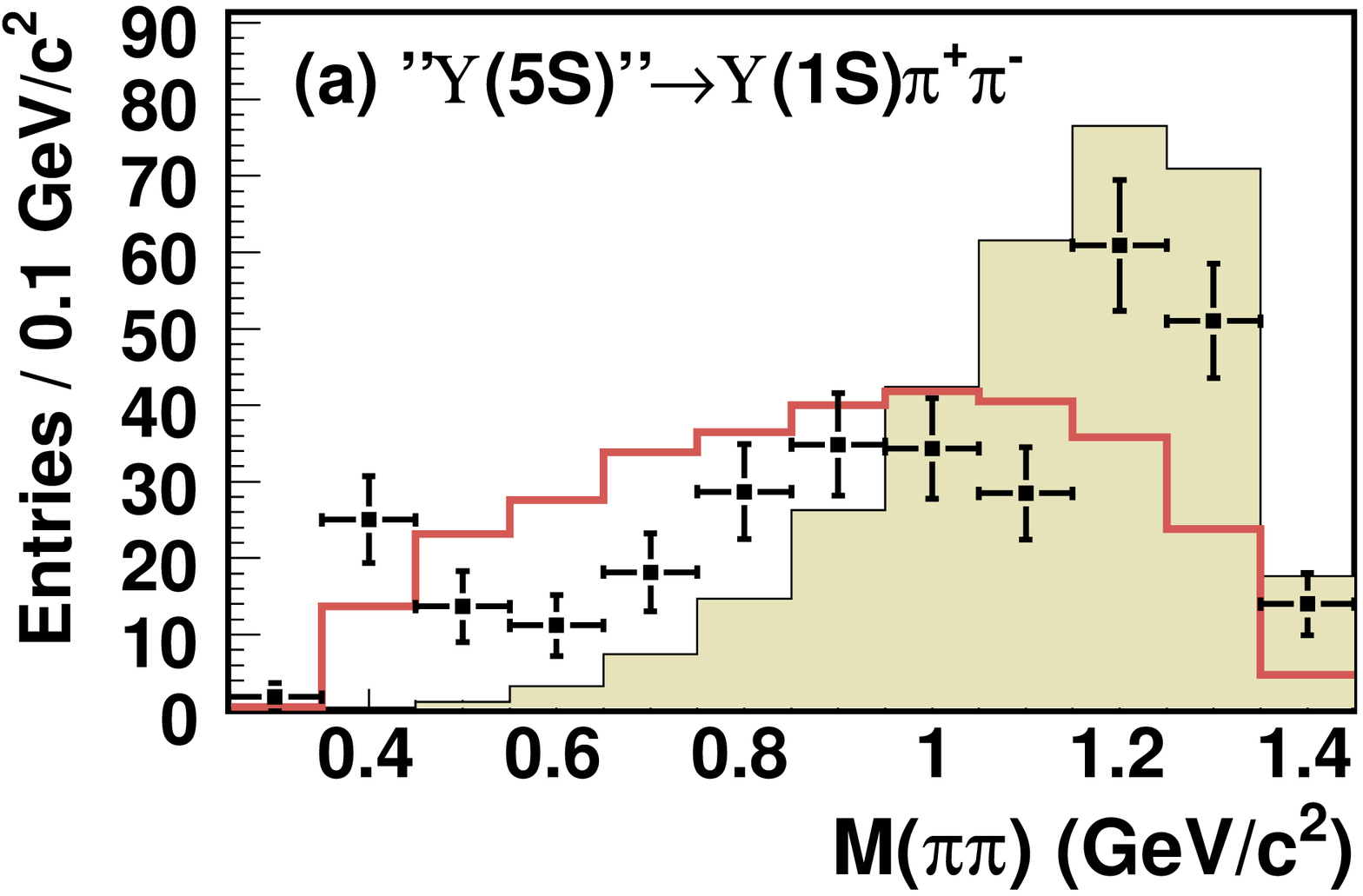}
\includegraphics[width=4.25cm]{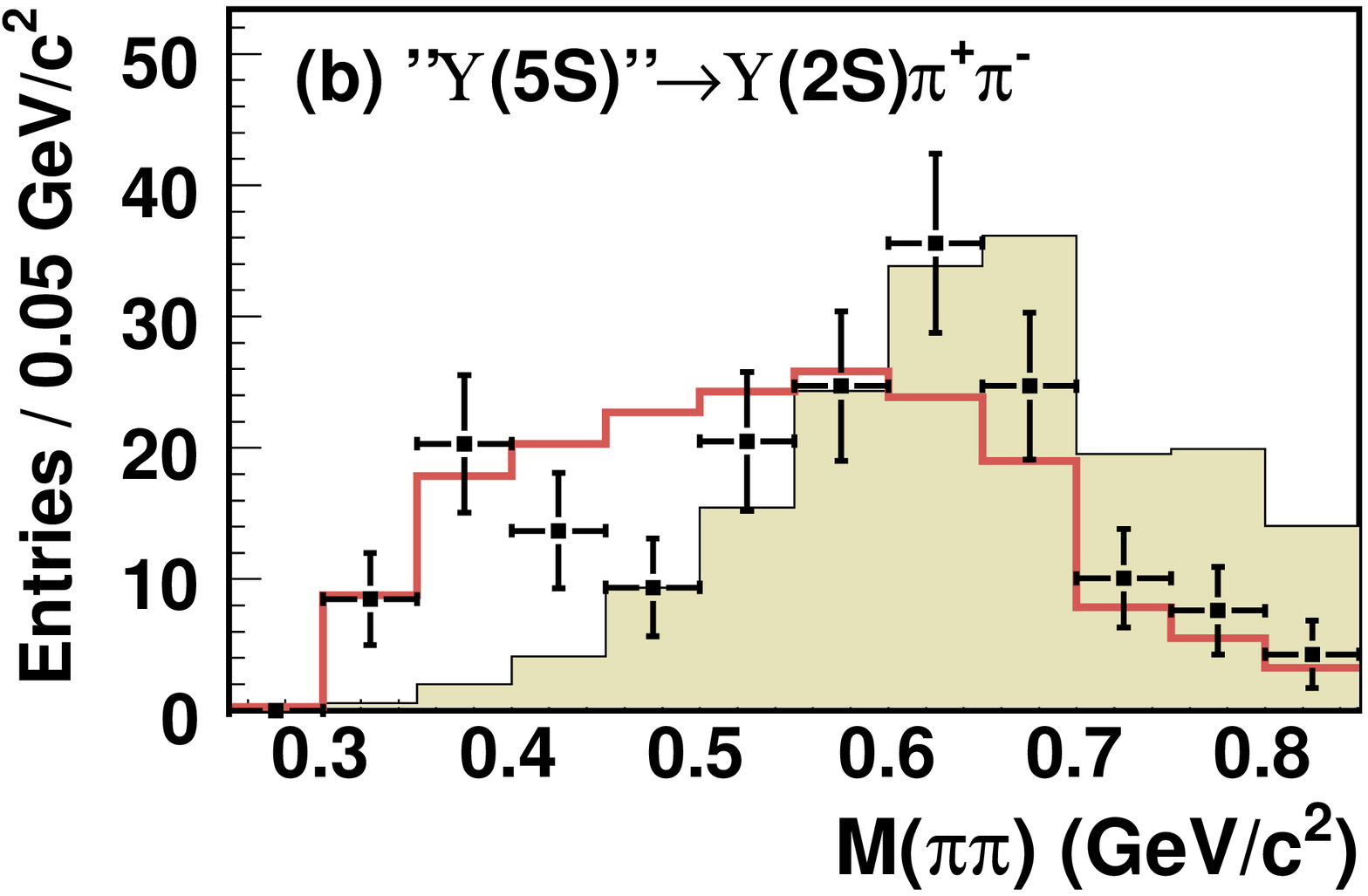}
\includegraphics[width=4.25cm]{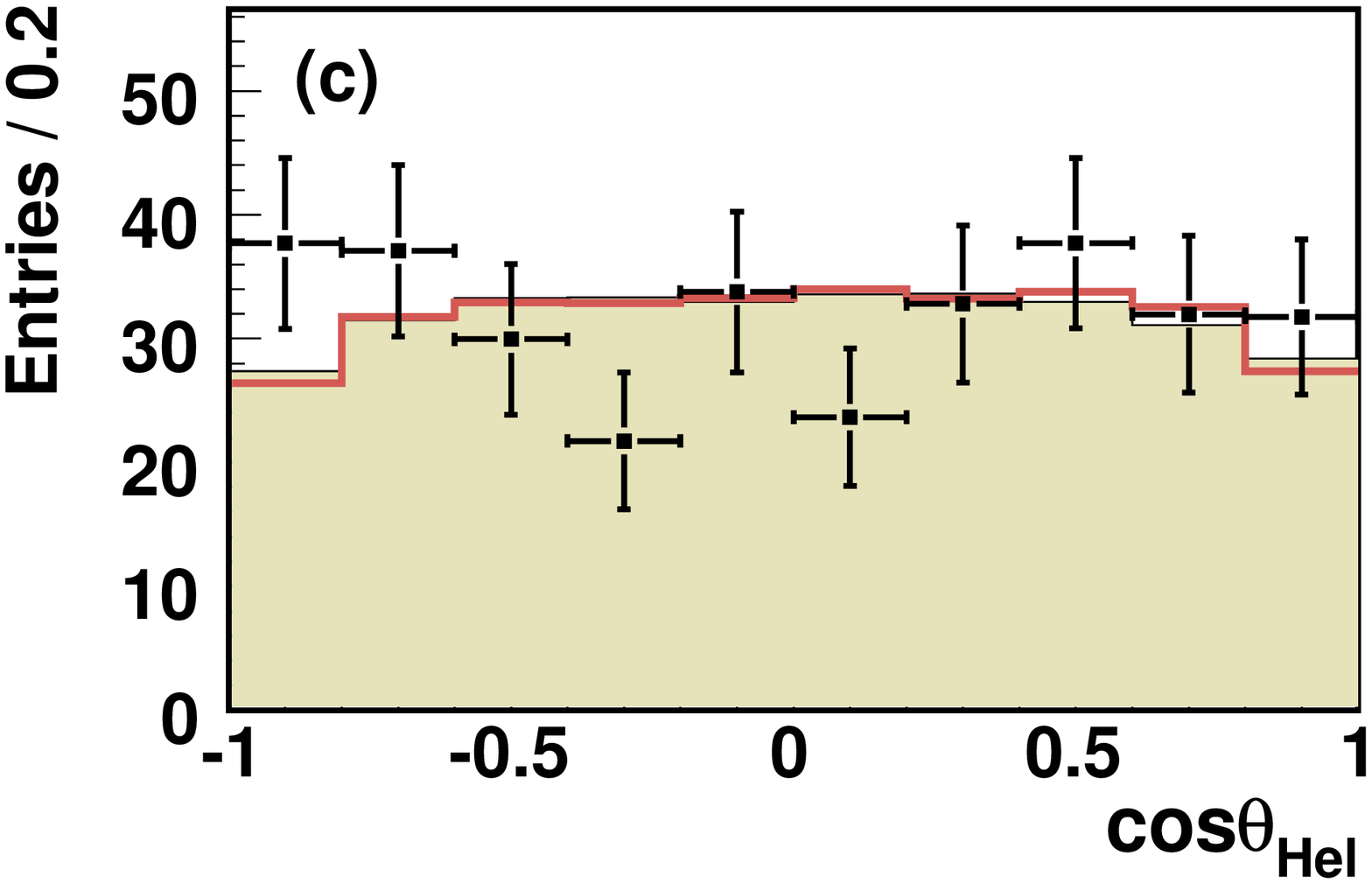}
\includegraphics[width=4.25cm]{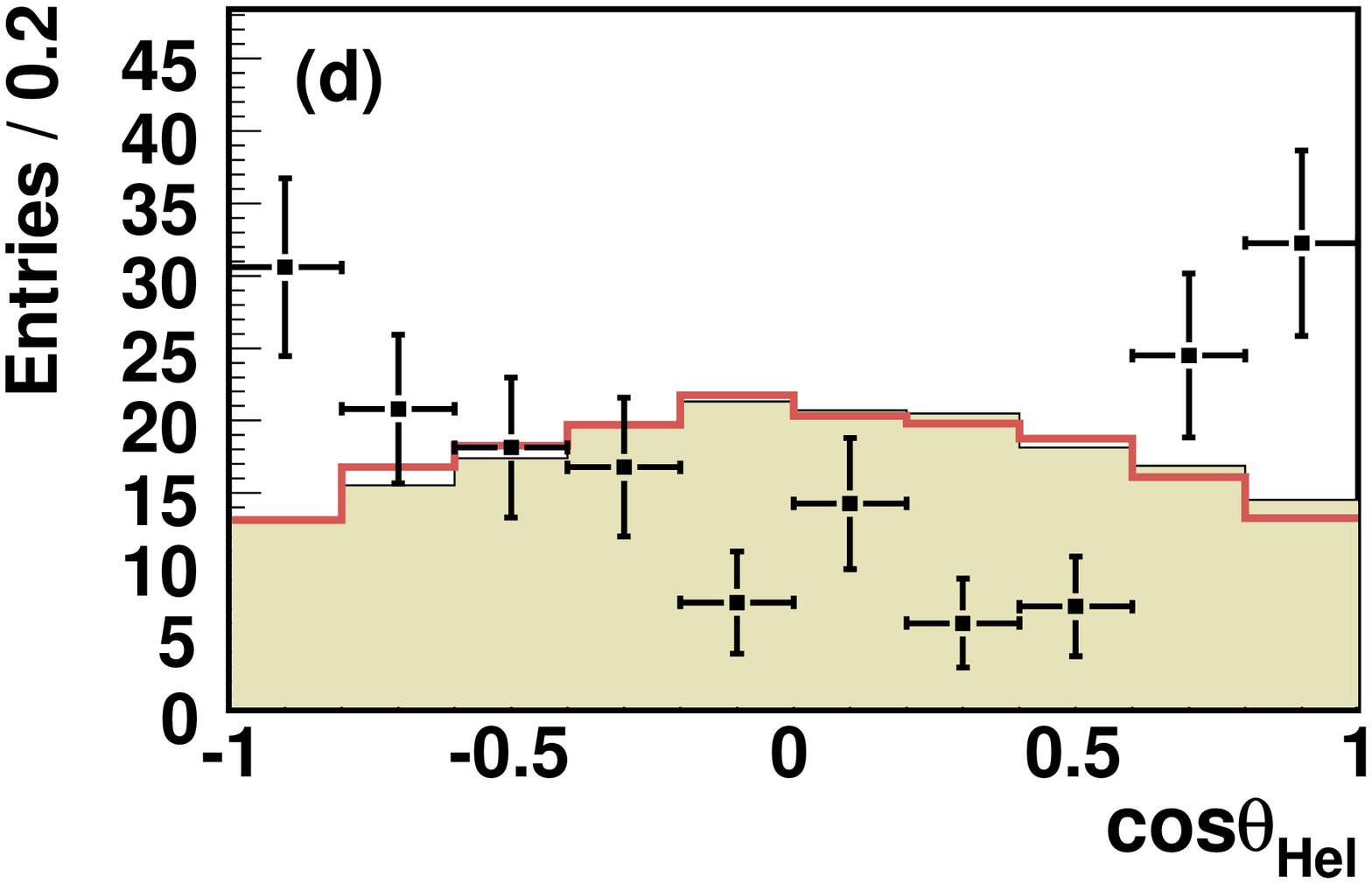}
\end{center}
\caption{The $\Upsilon(10860)$ yields as functions of $M(\pi^+\pi^-)$
and $\cos\theta_{\rm Hel}$ for (a,c) $\Upsilon(1S)\pi^+\pi^-$ and
(b,d) $\Upsilon(2S)\pi^+\pi^-$ transitions. 
The shaded (open) histogram
are from MC simulations using the model of
Ref.~\cite{Brown:1975dz} (phase space model). 
The numerical yields are given in the appendix.
} \label{fig:mpipi}
\end{figure}

The yields for $\Upsilon(10860)\to\Upsilon(1S)\pi^+\pi^-$,
$\Upsilon(2S)\pi^+\pi^-$ are found to be large; thus, the
corresponding invariant masses of the $\pi^+\pi^-$ system,
$M(\pi^+\pi^-)$, and the cosine of the helicity angle,
$\cos\theta_{\rm Hel}$, can be examined in detail. The helicity angle,
$\theta_{\rm Hel}$, is the angle between the $\pi^-$ and $\Upsilon(10860)$
momenta in the $\pi^+\pi^-$ rest frame.
Figure~\ref{fig:mpipi} shows the $\Upsilon(10860)$ yields as 
functions of $M(\pi^+\pi^-)$ and $\cos\theta_{\rm Hel}$, which 
are extracted using ML fits to $\Delta M$ in bins of  
$M(\pi^+\pi^-)$ or $\cos\theta_{\rm Hel}$. 
The shaded histograms in the figure are
the distributions from MC simulations using the
model of Ref.~\cite{Brown:1975dz}, while the open histograms
show a  generic phase space model. 
As neither model agrees well with the observed distributions 
and the efficiencies are sensitive to both variables, 
the reconstruction
efficiencies for $\Upsilon(10860)\to\Upsilon(1S)\pi^+\pi^-$ and
$\Upsilon(2S)\pi^+\pi^-$ are obtained using  MC
samples reweighted according to the measured $M(\pi^+\pi^-)$ and
$\cos\theta_{\rm Hel}$ spectra. Due to limited statistics, we
estimate the reconstruction efficiencies for
$\Upsilon(10860)\to\Upsilon(3S)\pi^+\pi^-$ and
$\Upsilon(1S)K^+K^-$ modes using the model of
Ref.~\cite{Brown:1975dz}.
Comparison of the $M(\pi^+ \pi^-)$ distribution obtained here 
with other $\Upsilon(nS)\to \Upsilon(mS) \pi^+ \pi^-$ ($m<n$) decays 
could be important for the theoretical interpretation of the results~\cite{Brown:1975dz,Simonov:2007bm}.

Assuming that signal events come only from the $\Upsilon(5S)$
resonance, the corresponding branching fractions and partial
widths can be extracted using ratios to the $\Upsilon(5S)$
cross section at $\sqrt{s}\sim10.87$ GeV, 
$0.302\pm0.015$ nb~\cite{Drutskoy:2006fg}.
The results, including the observed cross sections, are given in
Table~\ref{tab:results}. The values include the world average
branching fractions for $\Upsilon(nS)\to\mu^+\mu^-$ decays, and the total width of
the $\Upsilon(5S)$~\cite{ref:PDG2006}.  The measured partial widths, of order
$0.6$--$0.8$ MeV, are large compared to all other known 
transitions among $\Upsilon(nS)$ states. 
The partial widths for $\Upsilon(2S)$, $\Upsilon(3S)$, and
$\Upsilon(4S) \to \Upsilon(1S)\pi^+\pi^-$ transitions are all at
the keV level (Table~\ref{tab:bb_states}).

\begin{table*}[htbp]
\caption{Signal yield ($N_s$), significance ($\Sigma$),
reconstruction efficiency, and observed cross section ($\sigma$)
for $e^+e^- \to \Upsilon(nS)\pi^+\pi^-$ and $\Upsilon(1S)K^+K^-$
at $\sqrt{s} \sim 10.87$ GeV. Assuming the $\Upsilon(5S)$ to be the
sole source of the observed events, the branching fractions
($\mathcal{B}$) and the partial widths ($\Gamma$) for
$\Upsilon(5S) \to \Upsilon(nS)\pi^+\pi^-$ and $\Upsilon(1S)K^+K^-$
are also given. The first uncertainty is statistical, and the
second is systematic.} \label{tab:results}
\begin{center}
\begin{tabular}{lcccccccc}
\hline
\hline
Process & $N_s$ & $\Sigma$ & Eff.(\%) & $\sigma$(pb) &  $\mathcal{B}$(\%) & $\Gamma$(MeV) \\
\hline
$\Upsilon(1S)\pi^+\pi^-$~~ & $325^{+20}_{-19}$    &  20$\sigma$ & $37.4$  & ~~$1.61\pm0.10\pm0.12$              &  ~~$0.53\pm0.03\pm0.05$              & ~~$0.59\pm0.04\pm0.09$ \\
$\Upsilon(2S)\pi^+\pi^-$~~ & $186\pm15$           &  14$\sigma$ & $18.9$  & ~~$2.35\pm0.19\pm0.32$              &  ~~$0.78\pm0.06\pm0.11$              & ~~$0.85\pm0.07\pm0.16$ \\
$\Upsilon(3S)\pi^+\pi^-$~~ & $10.5^{+4.0}_{-3.3}$ & 3.2$\sigma$ &  $1.5$  & ~~$1.44^{+0.55}_{-0.45}\pm0.19$     &  ~~$0.48^{+0.18}_{-0.15}\pm0.07$     & ~~$0.52^{+0.20}_{-0.17}\pm0.10$ \\
$\Upsilon(1S)K^+K^-$~~     & $20.2^{+5.2}_{-4.5}$ & 4.9$\sigma$ & $20.3$  & ~~$0.185^{+0.048}_{-0.041}\pm0.028$ &  ~~$0.061^{+0.016}_{-0.014}\pm0.010$ & ~~$0.067^{+0.017}_{-0.015}\pm0.013$ \\
\hline
\hline
\end{tabular}
\end{center}
\end{table*}

\begin{table}[htpb]
\caption{The total width $\Gamma_{\rm total}$, and the partial width $\Gamma_{e^+e^-}$, 
$\Gamma_{\Upsilon(1S)\pi^+\pi^-}$.
Most values are from Refs.~\cite{ref:PDG2006,Aubert:2006bm,Sokolov:2006sd}.}
\label{tab:bb_states}
\begin{center}
\begin{tabular}{cccc}
\hline
\hline
Process & $\Gamma_{\rm total}$ & $\Gamma_{e^+e^-}$ & $\Gamma_{\Upsilon(1S)\pi^+\pi^-}$ \\
\hline
$\Upsilon(2S)\to\Upsilon(1S)\pi^+\pi^-$    & 0.032 MeV &  0.612 keV & 0.0060 MeV  \\
$\Upsilon(3S)\to\Upsilon(1S)\pi^+\pi^-$    & 0.020 MeV &  0.443 keV & 0.0009 MeV  \\
$\Upsilon(4S)\to\Upsilon(1S)\pi^+\pi^-$    & 20.5 MeV  &  0.272 keV & 0.0019 MeV  \\
$\Upsilon(10860)\to\Upsilon(1S)\pi^+\pi^-$ & 110 MeV   &  0.31 keV  & 0.59 MeV    \\
\hline
\hline
\end{tabular}
\end{center}
\end{table}

The systematic uncertainties for the cross sections are
dominated by the $\Upsilon(nS)\to\mu^+\mu^-$ branching fractions,
MC reconstruction efficiencies, and PDF parameterization for the
fits. 
Uncertainties of 2.0\%,
8.8\%, and 9.6\% for the $\Upsilon(1S)$, $\Upsilon(2S)$, and
$\Upsilon(3S)\to\mu^+\mu^-$ branching fractions are included,
respectively. 
For the $\Upsilon(1S)\pi^+\pi^-$ and $\Upsilon(2S)\pi^+\pi^-$ modes,
the reconstruction efficiencies are obtained from MC simulations
using the observed $M(\pi^+\pi^-)$ and $\cos\theta_{\rm Hel}$
distributions as inputs. 
The uncertainties associated with these
distributions give rise to 4.4\% and 6.8\% errors for the
$\Upsilon(1S)\pi^+\pi^-$ and $\Upsilon(2S)\pi^+\pi^-$ MC
efficiencies, respectively. 
For the other two channels, 
we try as input the models of Ref.~\cite{Brown:1975dz} and phase space model;
the corresponding variations in acceptance are included as 
systematic uncertainties.
A relatively large uncertainty of
13.6\% arises for the $\Upsilon(10860) \to\Upsilon(1S)K^+K^-$
channel, while the corresponding error for $\Upsilon(10860)
\to\Upsilon(3S)\pi^+\pi^-$ is small (3.2\%) due to the limited
phase space. The uncertainties from PDF parameterization are
obtained either by replacing the signal PDF with a sum of three
Gaussians, or by replacing the background PDF with a second-order polynomial. The
differences between the fit results obtained with alternative PDFs and the nominal
results are taken as the systematic uncertainty. The
selection criteria for rejecting
radiative Bhabha and $\mu$-pair events are examined using
the data~\cite{Tajima} collected at the $\Upsilon(3S)$ resonance.
The 1.9\% difference between data and MC efficiencies for
$\Upsilon(3S) \to \Upsilon(1S)\pi^+\pi^-$ decays is included as a
systematic uncertainty. Other uncertainties included are:
tracking efficiency ($1\%$ per charged track), muon
identification (0.5\% per muon candidate), electron rejection for
the charged pions (0.1--0.2\% per pion), kaon identification
(1.8\% per kaon), trigger efficiencies (0.9--4.5\%), and KEKB
luminosity (1.4\%). The uncertainties from all sources are added
in quadrature. The total systematic uncertainties are
7.5\%, 13.5\%, 13.1\%, and 15.3\% for the $\Upsilon(1S)\pi^+\pi^-$,
$\Upsilon(2S)\pi^+\pi^-$, $\Upsilon(3S)\pi^+\pi^-$, and
$\Upsilon(1S)K^+K^-$ channels, respectively.

For branching fraction estimation, the error in the 
$\Upsilon(5S)$ cross section ($\pm0.015$ nb) gives a 5.0\% uncertainty in
signal normalization. For the partial widths, there is an
additional uncertainty of 11.8\% coming from using the total width of the
$\Upsilon(5S)$.

In summary, we report the first observation of $e^+e^- \to
\Upsilon(1S)\pi^+\pi^-$ and $\Upsilon(2S)\pi^+\pi^-$ transitions, and first
evidence of $e^+e^- \to \Upsilon(3S)\pi^+\pi^-$ and
$\Upsilon(1S)K^+K^-$ transitions at a CM energy near the $\Upsilon(5S)$ resonance of
$\sqrt{s}\sim10.87$ GeV. Clear signals are observed at the expected
CM energy, with subsequent $\Upsilon(nS) \to \mu^+\mu^-$
decay. 
The measured cross sections are 
$1.61\pm0.10\pm0.12$ pb,		 
$2.35\pm0.19\pm0.32$ pb,	 
$1.44^{+0.55}_{-0.45}\pm0.19$ pb, and
$0.185^{+0.048}_{-0.041}\pm0.028$ pb for 
$e^+e^- \to \Upsilon(1S)\pi^+\pi^-$,
$\Upsilon(2S)\pi^+\pi^-$, $\Upsilon(3S)\pi^+\pi^-$, and
$\Upsilon(1S)K^+K^-$ transitions, respectively.
The first uncertainty is statistical, and the
second is systematic.
Assuming the observed signal events are due solely to the
$\Upsilon(5S)$ resonance, branching fractions are measured to be
in the range (0.48--0.78)\% for $\Upsilon(nS)\pi^+\pi^-$ channels, and 0.061\% for the $\Upsilon(1S)K^+K^-$ channel.
The corresponding
partial widths are found to be in the range (0.52--0.85) MeV for
$\Upsilon(nS)\pi^+\pi^-$, and 0.067 MeV for the $\Upsilon(1S)K^+K^-$
mode, more than two orders of magnitude
larger than the corresponding partial widths for
$\Upsilon(4S)$, $\Upsilon(3S)$ or $\Upsilon(2S)$ decays.
The unexpectedly large partial widths disagree with the
expectation for a pure $b\overline{b}$ state, unless there is a new
mechanism to enhance the decay rate. A detailed energy scan within
the $\Upsilon(5S)$ energy region would help to extract the resonant
spectrum; a comparison between the yield of
$\Upsilon(nS)\pi^+\pi^-$ events and the total hadronic
cross section may help us to understand the nature of the
signal.

We thank the KEKB group for excellent operation of the
accelerator, the KEK cryogenics group for efficient solenoid
operations, and the KEK computer group and
the NII for valuable computing and Super-SINET network
support.  We acknowledge support from MEXT and JSPS (Japan);
ARC and DEST (Australia); NSFC (China);
DST (India); MOEHRD, KOSEF, KRF and SBS Foundation (Korea);
KBN (Poland); MES and RFAAE (Russia); ARRS (Slovenia); SNSF (Switzerland);
NSC and MOE (Taiwan); and DOE (USA).

\appendix
\section{Appendix: Yields of $\Upsilon(10860) \to \Upsilon(1S)\pi^+\pi^-$ and 
$\Upsilon(2S)\pi^+\pi^-$ transitions as functions of 
$M(\pi^+\pi^-)$ and $\cos\theta_{\rm Hel}$}

\begin{table}[htpb]
\caption{Numerical yields of $\Upsilon(10860) \to \Upsilon(1S)\pi^+\pi^-$ and 
$\Upsilon(2S)\pi^+\pi^-$ transitions as functions of 
$M(\pi^+\pi^-)$ and $\cos\theta_{\rm Hel}$ which are 
shown in Figure~\ref{fig:mpipi}. The uncertainties of the yields are statistical only.}
\label{tab:num_table_4}
\begin{center}
\begin{tabular}{r|cr|cr}
\hline
\hline
    & \multicolumn{2}{c|}{$\Upsilon(1S)\pi^+\pi^-$} & \multicolumn{2}{c}{$\Upsilon(2S)\pi^+\pi^-$} \\
Bin & $M(\pi^+\pi^-)$ (GeV) & Yield & $M(\pi^+\pi^-)$ (GeV) & Yield \\
\hline
 1  &  $[0.25,0.35)$ &  $ 1.83_{-1.16}^{+1.84}$ &  $[0.25,0.30)$ &  $ 0.0 $		    \\
 2  &  $[0.35,0.45)$ &  $25.03_{-4.97}^{+5.70}$ &  $[0.30,0.35)$ &  $ 8.48_{-2.82}^{+3.51}$ \\
 3  &  $[0.45,0.55)$ &  $13.67_{-3.94}^{+4.66}$ &  $[0.35,0.40)$ &  $20.31_{-4.52}^{+5.22}$ \\
 4  &  $[0.55,0.65)$ &  $11.21_{-3.28}^{+3.99}$ &  $[0.40,0.45)$ &  $13.69_{-3.70}^{+4.38}$ \\
 5  &  $[0.65,0.75)$ &  $18.16_{-4.41}^{+5.12}$ &  $[0.45,0.50)$ &  $ 9.37_{-3.02}^{+3.73}$ \\ 
 6  &  $[0.75,0.85)$ &  $28.70_{-5.47}^{+6.18}$ &  $[0.50,0.55)$ &  $20.51_{-4.58}^{+5.27}$ \\
 7  &  $[0.85,0.95)$ &  $34.84_{-5.97}^{+6.70}$ &  $[0.55,0.60)$ &  $24.70_{-4.97}^{+5.70}$ \\
 8  &  $[0.95,1.05)$ &  $34.35_{-5.86}^{+6.57}$ &  $[0.60,0.65)$ &  $35.58_{-6.10}^{+6.83}$ \\
 9  &  $[1.05,1.15)$ &  $28.49_{-5.35}^{+6.03}$ &  $[0.65,0.70)$ &  $24.70_{-4.88}^{+5.58}$ \\
10  &  $[1.15,1.25)$ &  $60.91_{-7.87}^{+8.57}$ &  $[0.70,0.75)$ &  $10.07_{-3.03}^{+3.74}$ \\
11  &  $[1.25,1.35)$ &  $50.99_{-6.81}^{+7.48}$ &  $[0.75,0.80)$ &  $ 7.60_{-2.64}^{+3.35}$ \\
12  &  $[1.35,1.45)$ &  $14.00_{-3.41}^{+4.09}$ &  $[0.80,0.85)$ &  $ 4.28_{-1.88}^{+2.58}$ \\
\hline
Bin & $\cos\theta_{\rm Hel}$ & Yield & $\cos\theta_{\rm Hel}$ & Yield \\
\hline
 1  &  $[-1.0,-0.8)$ &  $37.68_{-6.24}^{+6.90}$ &  $[-1.0,-0.8)$ &  $30.59_{-5.46}^{+6.15}$ \\
 2  &  $[-0.8,-0.6)$ &  $37.09_{-6.20}^{+6.90}$ &  $[-0.8,-0.6)$ &  $20.82_{-4.49}^{+5.14}$ \\
 3  &  $[-0.6,-0.4)$ &  $29.95_{-5.41}^{+6.07}$ &  $[-0.6,-0.4)$ &  $18.15_{-4.14}^{+4.83}$ \\
 4  &  $[-0.4,-0.2)$ &  $21.74_{-4.84}^{+5.51}$ &  $[-0.4,-0.2)$ &  $16.78_{-4.07}^{+4.78}$ \\
 5  &  $[-0.2,+0.0)$ &  $33.77_{-5.85}^{+6.48}$ &  $[-0.2,+0.0)$ &  $ 7.45_{-2.78}^{+3.53}$ \\ 
 6  &  $[+0.0,+0.2)$ &  $23.65_{-4.84}^{+5.55}$ &  $[+0.0,+0.2)$ &  $14.28_{-3.76}^{+4.53}$ \\
 7  &  $[+0.2,+0.4)$ &  $32.80_{-5.67}^{+6.34}$ &  $[+0.2,+0.4)$ &  $ 6.00_{-2.36}^{+3.07}$ \\
 8  &  $[+0.4,+0.6)$ &  $37.72_{-6.17}^{+6.86}$ &  $[+0.4,+0.6)$ &  $ 7.15_{-2.75}^{+3.44}$ \\
 9  &  $[+0.6,+0.8)$ &  $31.97_{-5.65}^{+6.33}$ &  $[+0.6,+0.8)$ &  $24.49_{-4.95}^{+5.65}$ \\
10  &  $[+0.8,+1.0)$ &  $31.75_{-5.58}^{+6.28}$ &  $[+0.8,+1.0)$ &  $32.26_{-5.65}^{+6.41}$ \\
\hline
\hline
\end{tabular}
\end{center}
\end{table}

\end{document}